\documentclass[submission,creativecommons]{eptcs}

\providecommand{\event}{HCVS 2020} 
\usepackage{breakurl}             
\usepackage{underscore}           
\usepackage{mathtools} 
\usepackage{todonotes}
\usepackage{listings}
\usepackage{makecell}
\usepackage{xspace}
\usepackage{longtable,tablefootnote,tabularx}
\usepackage[title]{appendix}
\usepackage{color, colortbl}

\title{Competition Report: \CHCC
}
\author{Philipp R\"ummer
\institute{Uppsala University, Sweden}
}
\def\titlerunning{Competition Report: \CHCC}
\def\authorrunning{Philipp R\"ummer}

\newcommand{\LIA}{LIA-nonlin\xspace}
\newcommand{\LIAlin}{LIA-lin\xspace}
\newcommand{\LIAlinar}{LIA-lin-arrays\xspace}
\newcommand{\LRATS}{LRA-TS\xspace}

\newcommand{\CHC}{CHC-COMP\xspace}
\newcommand{\CHCC}{CHC-COMP-20\xspace}
\newcommand{\CHCCold}{CHC-COMP-19\xspace}
\newcommand{\CHCColdold}{CHC-COMP-18\xspace}
\newcommand{\CHCCnew}{CHC-COMP-21\xspace}

\newcommand{\Score}{\textbf{Score}\xspace}
\newcommand{\CPUtime}{\textbf{CPU time}\xspace}
\newcommand{\Walltime}{\textbf{Wall-clock time}\xspace}
\newcommand{\Speedup}{\textbf{Speedup}\xspace}
\newcommand{\SotAC}{\textbf{SotAC}\xspace}
\newcommand{\sat}{\textbf{sat}\xspace}
\newcommand{\unsat}{\textbf{unsat}\xspace}
\newcommand{\unknown}{\textbf{unknown}\xspace}

\newcommand{\StarExec}{\href{https://www.starexec.org}{StarExec}\xspace}

\begin{document}
\maketitle

\begin{abstract}
  \CHCC\footnote{\url{https://chc-comp.github.io/}} is the third
  competition of solvers for Constrained Horn Clauses. In this year, 9
  solvers participated at the competition, and were evaluated in four
  separate tracks on problems in linear integer arithmetic, linear
  real arithmetic, and arrays. The competition was run in the first
  week of May 2020 using the \StarExec computing cluster. This report
  gives an overview of the competition design, explains the
  organisation of the competition, and presents the competition
  results.
\end{abstract}

\section{Introduction}

Constrained Horn Clauses (CHC) have over the last decade emerged as a
uniform framework for reasoning about different aspects of software
safety~\cite{andrey-pldi,BjornerGMR15}. Constrained Horn clauses form
a fragment of first-order logic, modulo various background theories,
in which models can be constructed effectively with the help of model
checking algorithms. Horn clauses can be used as an intermediate
verification language that elegantly captures various classes of
systems (e.g., sequential code, programs with functions and
procedures, concurrent programs, or networks of timed automata) and
various verification methodologies (e.g., the use of state invariants,
verification with the help of contracts, Owicki-Gries-style
invariants, or rely-guarantee methods). Horn solvers can be used as
off-the-shelf back-ends in verifiers, and thus enable construction of
verification systems in a modular way.

\CHCC is the third competition of solvers for Constrained Horn
Clauses, a competition affiliated with the 7th Workshop on Horn
Clauses for Verification and Synthesis (HCVS) at ETAPS~2020.  The goal
of \CHC is to compare state-of-the-art tools for Horn solving with
respect to performance and effectiveness on realistic, publicly
available benchmarks.  The deadline for submitting solvers to \CHCC
was April~30 2020, resulting in 9 solvers participating, which were
evaluated in the first week of May~2020. The 9 solvers were evaluated
in four separate tracks on problems in linear integer arithmetic,
linear real arithmetic, and the theory of arrays. The results of the
competition can be found in Section~\ref{sec:results} of this report.

Due to the Covid-19 crisis, both ETAPS~2020 and HCVS were postponed,
and at the time of finalising this report no new dates had been set;
this means that the present report is the main documentation of \CHCC
at the moment. It is planned, however, that the competition will be
presented and discussed in detail also at the HCVS workshop when it
takes place, either in physical or virtual form.

\subsection{Acknowledgements}

\CHCC heavily builds on the infrastructure and scripts written for
\CHCColdold and \CHCCold, run by Arie Gurfinkel and Grigory
Fedyukovich, respectively. Contributors to the competition
infrastructure also include Adrien Champion, Dejan Jovanovic, and
Nikolaj Bj\o rner.

Like in the first two competitions, \CHCC was run using
\StarExec \cite{DBLP:conf/cade/StumpST14}. We are extremely grateful
for the computing resources and evaluation environment provided by
\StarExec, and for the fast and competent support by Aaron Stump and
his team whenever problems occurred. \CHCC would not have been
possible without this!

The organiser of \CHCC is supported by the Swedish Research Council
(VR) under grant 2018-04727, and by the Swedish Foundation for
Strategic Research (SSF) under the project WebSec (Ref.\ RIT17-0011).

\section{Brief Overview of the Competition Design}

\subsection{Competition Tracks}

In \CHCC the same tracks as in \CHCCold were used for evaluating solvers:
\begin{itemize}
\item \textbf{\LIA}: benchmarks with at least one non-linear clause,
  and linear integer arithmetic as background theory;
\item \textbf{\LIAlin}: benchmarks with only linear clauses, and
  linear integer arithmetic as background theory;
\item \textbf{\LIAlinar}: benchmarks with only linear clauses, and the
  combined theory of linear integer arithmetic and arrays as
  background theory;
\item \textbf{\LRATS}: benchmarks encoding transition systems, with
  linear real arithmetic as background theory. Benchmarks in this
  track have exactly one uninterpreted relation symbol, and exactly
  three linear clauses encoding initial states, transitions, and error
  states.
\end{itemize}

\subsection{Computing Nodes}

The competition was run on 30 nodes provided by
\StarExec. Each node had two quadcore
CPUs, and each node was used to run two jobs in parallel during the
competition runs.
The
\href{https://www.starexec.org/starexec/public/machine-specs.txt}{machine
  specs} are:
\begin{verbatim}
Intel(R) Xeon(R) CPU E5-2609 0 @ 2.40GHz (2393 MHZ)
    10240  KB Cache
    263932744 kB main memory

# Software:
OS:       CentOS Linux release 7.7.1908 (Core)
kernel:   3.10.0-1062.4.3.el7.x86_64
glibc:    glibc-2.17-292.el7.x86_64
          gcc-4.8.5-39.el7.x86_64
          glibc-2.17-292.el7.i686
\end{verbatim}

\subsection{Test and Competition Runs}

All solvers submitted to \CHCC were evaluated twice:
\begin{itemize}
\item in a first set of \textbf{test runs}, in which pre-submissions
  of the solvers were evaluated to check their configurations and
  identify possible inconsistencies. For the test runs a smaller set
  of randomly selected benchmarks was used.  The results of the test
  runs for each solver were directly communicated to the team
  submitting the solver, but not made public and not shared with other
  teams.  In the test runs, each solver-benchmark pair was limited to
  600s CPU time, 600s wall-clock time, and 64GB memory.
\item in the \textbf{competition runs}, the results of which
  determined the outcome of \CHCC. The selection of the benchmarks for
  the competition runs is described in Section~\ref{sec:selection},
  and the evaluation of the competition runs in
  Section~\ref{sec:eval}. In the competition runs, each job was
  limited to 1800s CPU time, 1800s wall-clock time, and 64GB memory.
\end{itemize}

\subsection{Evaluation of the Competition Runs}
\label{sec:eval}

The ranking of solvers in each track was done based on the \Score
reached by the solvers in the competition run for that track.  In case
two solvers had equal \Score, the ranking of the two solvers was
determined by \CPUtime. It was assumed that the outcome
of running one solver on one benchmark can only be \sat, \unsat, or
\unknown; the last outcome includes solvers giving up, running out of
resources, or crashing.

\medskip
The definition of \Score and \CPUtime are:
\begin{itemize}
\item \Score: the number of \sat or \unsat results produced by a
  solver on the benchmarks of a track.
\item \CPUtime: the average CPU time needed by a solver to produce the
  result \sat or \unsat in some track, not counting the runtime for
  \unknown results. This average time was computed based on the CPU
  time stamp reported by \StarExec for the output of the result
  \texttt{sat}/\texttt{unsat}.
\end{itemize}

In addition, the following features were computed for each solver and
each track:
\begin{itemize}
\item \Walltime: the average wall-clock time needed by a solver to
  produce the result \sat or \unsat in some track, not counting the
  runtime for \unknown results.
\item \Speedup: the ratio \CPUtime{} / \Walltime.
\item \SotAC: the state-of-the-art contribution of a solver, computed
  in the same way as in the \href{http://www.tptp.org/CASC/}{CADE ATP
    System Competition (CASC)}. The \SotAC of a benchmark in some
  track is the inverse of the number of systems that reported \sat or
  \unsat for the benchmark. The \SotAC of a solver is the average
  \SotAC of the benchmarks it could solve.
\end{itemize}

\section{Competition Benchmarks}

\subsection{File Format}

\CHC represents benchmarks in a fragment of the SMT-LIB 2.6
format. The fragment is defined on
\url{https://chc-comp.github.io/format.html}. For \CHCC, several minor
modifications were done in the format definition, in particular the
role of nullary predicates was clarified. The conformance of a
well-typed SMT-LIB script with the \CHC fragment can be checked using
the \texttt{format-checker} available on
\url{https://github.com/chc-comp/chc-tools}.

\subsection{Benchmark Processing}
\label{sec:processing}

All benchmarks used in \CHCC were pre-processed using the
\texttt{format.py} script available in the repository
\url{https://github.com/chc-comp/scripts}, using the command line
\begin{verbatim}
> python3 format.py --out_dir <outdir> --merge_queries True <smt-file>
\end{verbatim}
The script tries to translate arbitrary Horn-like problems in SMT-LIB
format to problems within the \CHC fragment.  Only benchmarks
processed in this way were used in the competition.  The option
\verb!--merge_queries! was added for \CHCC to the script, and has the
effect of merging multiple queries in a benchmark into a single query
by introducing an auxiliary nullary predicate. For a discussing on
this, see Section~\ref{sec:multiQueries}.

After processing with \texttt{format.py}, benchmarks were checked and
categorised into the four tracks using the \texttt{format-checker}
scripts available on
\url{https://github.com/chc-comp/chc-tools}.

Benchmarks that could not be processed by \texttt{format.py}, were
rejected by the \texttt{format-checker}, or did not conform to any of
the competition tracks, were not used in \CHCC.

\subsection{Handling of Benchmarks with Multiple Queries}
\label{sec:multiQueries}

In the \CHC format, a query is a clause with the head
\texttt{false}, and encodes the property to be verified in a
benchmark. At the moment, the \CHC fragment of SMT-LIB only allows
benchmarks with exactly one query, which means that problems involving
multiple queries have to be mapped to the single-query format. Of the
\CHC benchmarks, a significant number contains multiple
queries.

In the past competitions, this was handled by splitting benchmarks
with multiple queries into multiple single-query benchmarks (option
\verb!--split_queries!  of \verb!format.py!); in \CHCC, instead
benchmarks with multiple queries were converted to the single-query
format by introducing an auxiliary nullary predicate (or Boolean
variable), and merging all queries to a single query
(\verb!--merge_queries!  of \verb!format.py!). The motivation for the
new pre-processing is that splitting of queries sometimes makes
benchmarks artificially hard: with multiple queries, often some of the
queries can be disproven easily, while other queries are difficult to
prove or disprove. Such problems, seen as a whole, are simple to
solve, but splitting and considering all queries individually will
lead to some hard benchmarks. Since the benchmarks used in \CHC should
represent realistic applications, this is an undesired effect.

In future editions of \CHC, it might be even better to just allow
benchmarks with multiple queries, and extend the \CHC format
accordingly. Many solvers are already now able to process benchmarks
with multiple queries. Other solvers could decide themselves whether
benchmarks with multiple queries should be handled by splitting or by
merging; both translations are quite simple to implement in a solver,
or alternatively a solver could invoke the existing script for
pre-processing.

\subsection{Benchmark Inventory}

\begin{table}[tb]
  \begin{center}
  \newcommand{\empt}{\multicolumn{2}{c}{}}
  \begin{tabular}{l*{4}{r@{ / }l}}
    \textbf{Repository} &
               \multicolumn{2}{c}{\LIA} &
               \multicolumn{2}{c}{\LIAlin} &
               \multicolumn{2}{c}{\LIAlinar} &
               \multicolumn{2}{c}{\LRATS}
    \\\hline\hline
    eldarica-misc & 69&66 & 147&134 
    \\\hline
    extra-small-lia & \empt & 55&55 
    \\\hline
    hcai-bench & 135&133 & 100&86 & 39&39 
    \\\hline
    hopv & 68&67 & 49&48 
    \\\hline
    jayhorn-benchmarks &  5138&5084 & 75&73 
    \\\hline
    kind2-chc-benchmarks & 851&738 
    \\\hline
    ldv-ant-med & \empt & \empt & 10&10
    \\\hline
    llreve-bench & 59&57 & 44&44 & 31&31 
    \\\hline
    quic3 & \empt & \empt &43&43 
    \\\hline
    sally-chc-benchmarks & \empt & \empt & \empt & 177 & 174
    \\\hline
    seahorn & 72&70 & 3421&2847 
    \\\hline
    tricera-benchmarks & 4&4 & 405&405 
    \\\hline
    vmt-chc-benchmarks & \empt & 906&803 & \empt & 99 & 98
    \\\hline\hline
    sv-comp & 1643&1169 & 3150&2932 & 79&73 
    \\\hline
    chc-comp19-benchmarks
                        & 271&265 & 326&314 & 305&305 & 229&227
    \\\hline\hline
    \textbf{Total} & ~~8310&7653~~ & ~~8678&7741~~ & ~~~~~507&501 & 505&499
  \end{tabular}
  \end{center}

  \caption{Summary of benchmarks available on
    \url{https://github.com/chc-comp} and in the
    \href{https://www.starexec.org/starexec/secure/explore/spaces.jsp?id=73700}{StarExec
      CHC space}.  For each collection of benchmarks and each \CHCC
    track, the first number gives the total number of benchmarks, and
    the second number the number of contributed unique benchmarks
    (after discarding duplicate benchmarks).}
  \label{tab:benchmarks}
\end{table}

In contrast to most other competitions, \CHC stores benchmarks in a
decentralised way, in multiple repositories managed by the
contributors of the benchmarks themselves.  Table~\ref{tab:benchmarks}
summarises the number of benchmarks that were obtained by collecting
benchmarks from all available repositories using the process in
Section~\ref{sec:processing}. Duplicate benchmarks were identified by
computing a checksum for each (processed) benchmark, and were discarded.

The repository chc-comp19-benchmarks of benchmarks selected for
\CHCCold was included in the collection, because this repository
contains several unique families of benchmarks that are not available
in other repositories under \url{https://github.com/chc-comp}. Such
benchmarks include problems generated by the Ultimate tools in the
\LIAlinar track.

From jayhorn-benchmarks, only the problems generated for sv-comp-2020
were considered, which subsume the problems for sv-comp-2019.

\section{Benchmark Rating and Selection}
\label{sec:selection}

\begin{table}[tb]
  \begin{center}
  \newcommand{\empt}{\multicolumn{3}{c}{}}
  \begin{tabular}{l*{2}{r@{ / }r@{ / }r}}
    &          \multicolumn{3}{c}{~~~~~~~\LIA} &
               \multicolumn{3}{c}{~~~~~~~~\LIAlin}
                                          \\
    \textbf{Repository} &
               \#A & \#B & \#C &
               \#A & \#B & \#C
    \\\hline\hline
    eldarica-misc & 12&28&26 & 26&91&17
    \\\hline
    extra-small-lia & \empt & 3&24&28
    \\\hline
    hcai-bench & 19&71&43 & 59&19&8
    \\\hline
    hopv & 26&38&3 & 45&2&1
    \\\hline
    jayhorn-benchmarks &  49&2680&2355 & 55&18
    \\\hline
    kind2-chc-benchmarks & 58&179&501 
    \\\hline
    llreve-bench & 6&35&16 & 9&35
    \\\hline
    seahorn & 6&34&30 & 753&323&1771
    \\\hline
    tricera-benchmarks & 1&3& & 9&23&373
    \\\hline
    vmt-chc-benchmarks & \empt & 33&252&518
    \\\hline\hline
    sv-comp & 25&1057&87 & 968&1855&109
    \\\hline
    chc-comp19-benchmarks & 42&116&107 & 31&100&183
    \\\hline\hline
    \textbf{Total} & ~~~~~244&4241&3168 & ~~~~~1991&2742&3008
  \end{tabular}
  \end{center}

  \caption{The number of unique \LIA and \LIAlin benchmarks with ratings
    A / B / C.}
  \label{tab:rating}
\end{table}

\begin{table}[tb]
  \begin{center}
  \newcommand{\empt}{\multicolumn{2}{c}{}}
  \begin{tabular}{l*{2}{@{~~~~~~~~~~~~}r@{~~~~}r}@{\qquad}*{2}{r}}
    &          \multicolumn{2}{l}{\hspace*{-2ex}\LIA} &
               \multicolumn{2}{l}{\LIAlin} &
               \LIAlinar & \LRATS
                                          \\
    \textbf{Repository} &
               $N_r$ & \#Sel & 
               $N_r$ & \#Sel & \#Selected~~~~ & \#Selected
    \\\hline\hline
    eldarica-misc & 10&30 & 15&45
    \\\hline
    extra-small-lia & \empt & 10&30
    \\\hline
    hcai-bench & 20&60 & 15&38 & 39~~~~~~~~
    \\\hline
    hopv & 10&23 & 10&13
    \\\hline
    jayhorn-benchmarks &  30&90 & 10&20
    \\\hline
    kind2-chc-benchmarks & 30&90
    \\\hline
    ldv-ant-med & \empt & \empt & 10~~~~~~~~
    \\\hline
    llreve-bench & 15&45 & 15&30 & 31~~~~~~~~
    \\\hline
    quic3 & \empt & \empt &43~~~~~~~~
    \\\hline
    sally-chc-benchmarks & \empt & \empt & & 174~~~~~
    \\\hline
    seahorn & 15&45 & 30&90
    \\\hline
    tricera-benchmarks & 1&2 & 20&60
    \\\hline
    vmt-chc-benchmarks & \empt & 30&90 && 98~~~~~
    \\\hline\hline
    sv-comp & 30&90 & 30&90 & 73~~~~~~~~
    \\\hline
    chc-comp19-benchmarks & 30&90 & 30&90 & 305~~~~~~~~ & 227~~~~~
    \\\hline\hline
    \textbf{Total} & &565 & &596 & 501~~~~~~~~ & 499~~~~~
  \end{tabular}
  \end{center}

  \caption{The number of selected unique benchmarks for the four \CHCC
    tracks.}
  \label{tab:selectionNum}
\end{table}

This section describes how the benchmarks for \CHCC were selected
among the unique benchmarks summarised in
Table~\ref{tab:benchmarks}. For the competition tracks \LIAlinar and
\LRATS, the benchmark library only contains 501 and 499 unique
benchmarks, respectively, which are small enough sets to use all
benchmarks in the competition. For the tracks \LIA and \LIAlin, in
contrast, too many benchmarks are available, so that a representative
sample of the benchmarks had to be chosen.

To gauge the difficulty of the available problems in \LIA and \LIAlin,
a simple rating based on the performance of the \CHCCold solvers was
computed. For this, the two top-ranked competing solvers from \CHCCold
were run for a few seconds on each of the \LIA and \LIAlin benchmarks:
\begin{itemize}
\item \textbf{Eldarica:} was run with a 5s timeout on each
  benchmark.\footnote{Run on an Intel Core~i5-650 2-core machine with
    3.2GHz.} Eldarica runs entirely in managed code on a JVM; to avoid
  frequent JVM restarts, the daemon mode of Eldarica was
  used. Otherwise, the same binary and same options were chosen as in
  \CHCCold.
\item \textbf{Ultimate Unihorn:} was run with a slightly
  higher timeout of 8s, since the solver also partly runs in managed
  code, but (to the best of the organiser's knowledge) does not have a
  similar daemon mode as Eldarica. The same binary and options as in
  \CHCCold were used.
\end{itemize}

The outcomes of those test runs gave rise to three possible ratings
for each benchmark:
\begin{itemize}
\item \textbf{A:} both tools were able to determine the benchmark
  status within the given time budget.
\item \textbf{B:} only one tool could determine the benchmark status.
\item \textbf{C:} both tools timed out.
\end{itemize}

The number of benchmarks per rating are shown in
Table~\ref{tab:rating}. As can be seen from the table, the simple
rating method separates the benchmarks into partitions of comparable
size, and provides some information about the relative hardness of the
problems in the different repositories.

From each repository~$r$, up to $3\cdot N_r$ benchmarks were then
selected randomly: $N_r$ benchmarks with rating~A, $N_r$ benchmarks
with rating~B, and $N_r$ benchmarks with rating~C. If a repository
contained fewer than $N_r$ benchmarks for some particular rating,
instead benchmarks with the next-higher rating were chosen. As special
cases, up to $N_r$ benchmarks were selected from repositories with
only A-rated benchmarks; up to $2\cdot N_r$ benchmarks from
repositories with only B-rated benchmarks; and up to $3\cdot N_r$
benchmarks from repositories with only C-rated benchmarks.

The number~$N_r$ was chosen individually for each repository, based on
a manual inspection of the repository to judge the diversity of the
contained benchmarks. The chosen~$N_r$, and the numbers of selected
benchmarks for each repository, are given in
Table~\ref{tab:selectionNum}.

For the actual selection of benchmarks with rating~X, the following
Unix command was used:
\begin{verbatim}
> cat <rating-X-benchmark-list> | sort -R | head -n <num>
\end{verbatim}

The final set of benchmarks selected for \CHCC can be found in the
github repository
\url{https://github.com/chc-comp/chc-comp20-benchmarks}, and on
\StarExec in the public space \linebreak
\href{https://www.starexec.org/starexec/secure/explore/spaces.jsp?id=405635#}{\texttt{CHC/CHC-COMP/chc-comp20-benchmarks}}.

\section{Solvers Entering \CHCC}

\begin{table}[tb]
  \begin{center}
  \begin{tabularx}{\linewidth}{X*{4}{X}}
    \textbf{Solver} &  \LIA & \LIAlin & \LIAlinar & \LRATS
    \\\hline\hline
    Eldarica-abs & \ttfamily def & \ttfamily def & --- & ---
    \\\hline
    IC3IA & --- & \ttfamily default.sh & \ttfamily default.sh & \ttfamily default.sh
    \\\hline
    PCSat & \ttfamily pcsat\_dt\_tb &  \ttfamily pcsat\_dt\_tb & --- & ---
    \\\hline
    ProphIC3 & --- & --- & \ttfamily default.sh & ---
    \\\hline
    Sally & --- & --- & --- &
    \ttfamily
    y2o2_decomposing_itp,
    parallel
    \\\hline
    Spacer
           & \ttfamily lia  
           & \ttfamily lia_lin  
           &\ttfamily  arrays  
           & \ttfamily lra
    \\\hline
    \raggedright
    Ultimate TreeAutomizer & \ttfamily default & \ttfamily default & \ttfamily default & \ttfamily default
    \\\hline
    \raggedright
    Ultimate Unihorn & \ttfamily default & \ttfamily default & \ttfamily default & \ttfamily default
    \\\hline\hline
    \raggedright
    Eldarica (Hors Concours) & \ttfamily def   & \ttfamily def & \ttfamily def & ---
    \\\hline
  \end{tabularx}
  \end{center}
  \caption{The submitted solvers, and the configurations used in the
  individual tracks.}
  \label{tab:solvers}
\end{table}

In total, 9~solvers were submitted to \CHCC: 8 competing solvers, and
one further solver (Eldarica, co-developed by the competition
organiser) that was entering outside of the competition.  A summary of
the participating solvers is given in Table~\ref{tab:solvers}.

More details about the participating solvers are provided in the
solver descriptions in Section~\ref{sec:solvers}. The binaries of the
solvers used for the competition runs can be found in the public
\StarExec space \linebreak
\href{https://www.starexec.org/starexec/secure/explore/spaces.jsp?id=405635#}{\texttt{CHC/CHC-COMP/chc-comp20-benchmarks}}.

\section{Competition Results}
\label{sec:results}

\subsection{Overview}

The winners and top-ranked solvers of the four \CHCC tracks are:
\begin{center}
  \begin{tabularx}{\linewidth-8ex}{l*{4}{X}}
    \hline
    & \LIA & \LIAlin & \LIAlinar & \LRATS
    \\\hline\hline
    \textbf{Winner}~~~ & \textbf{Spacer} & \textbf{Spacer} & \textbf{Spacer} & \textbf{IC3IA}
    \\\hline
    Place 2& Eldarica-abs & Eldarica-abs & \raggedright Ultimate Unihorn & Sally\newline (two config.)
    \\\hline
    Place 3 & \raggedright Ultimate Unihorn & \raggedright Ultimate Unihorn & ProphIC3 & Spacer
    \\\hline
  \end{tabularx}
\end{center}

\subsection{Detailed Results}

\begin{figure}[p]
  \begin{center}
    \begin{tabular}{l*{7}{r}}
      \hline
      \textbf{Solver} & Score & \#sat & \#unsat & CPU time (s) &
            Wall-clock (s) & Speedup & SotAC
      \\\hline\hline
Spacer & 554 & 292 & 262 & 6.03 & 6.11 & 0.99 & 0.28\\\hline
\cellcolor{lightgray}Eldarica (HC) &\cellcolor{lightgray}513 &\cellcolor{lightgray}265 &\cellcolor{lightgray}248 &\cellcolor{lightgray}43.58 &\cellcolor{lightgray}19.10 &\cellcolor{lightgray}2.28 &\cellcolor{lightgray}0.23\\\hline
Eldarica-abs & 513 & 266 & 247 & 52.07 & 35.96 & 1.45 & 0.23\\\hline
U. Unihorn & 420 & 212 & 208 & 75.73 & 49.11 & 1.54 & 0.21\\\hline
PCSat & 331 & 156 & 175 & 92.10 & 29.54 & 3.12 & 0.20\\\hline
U. TreeAutomizer & 118 & 34 & 84 & 41.17 & 30.00 & 1.37 & 0.17\\\hline\hline
      Any solver & 560 & 298 & 262\\\hline
    \end{tabular}
  \end{center}

  \bigskip
  \includegraphics[scale=0.62,trim=15 0 100 0,clip]{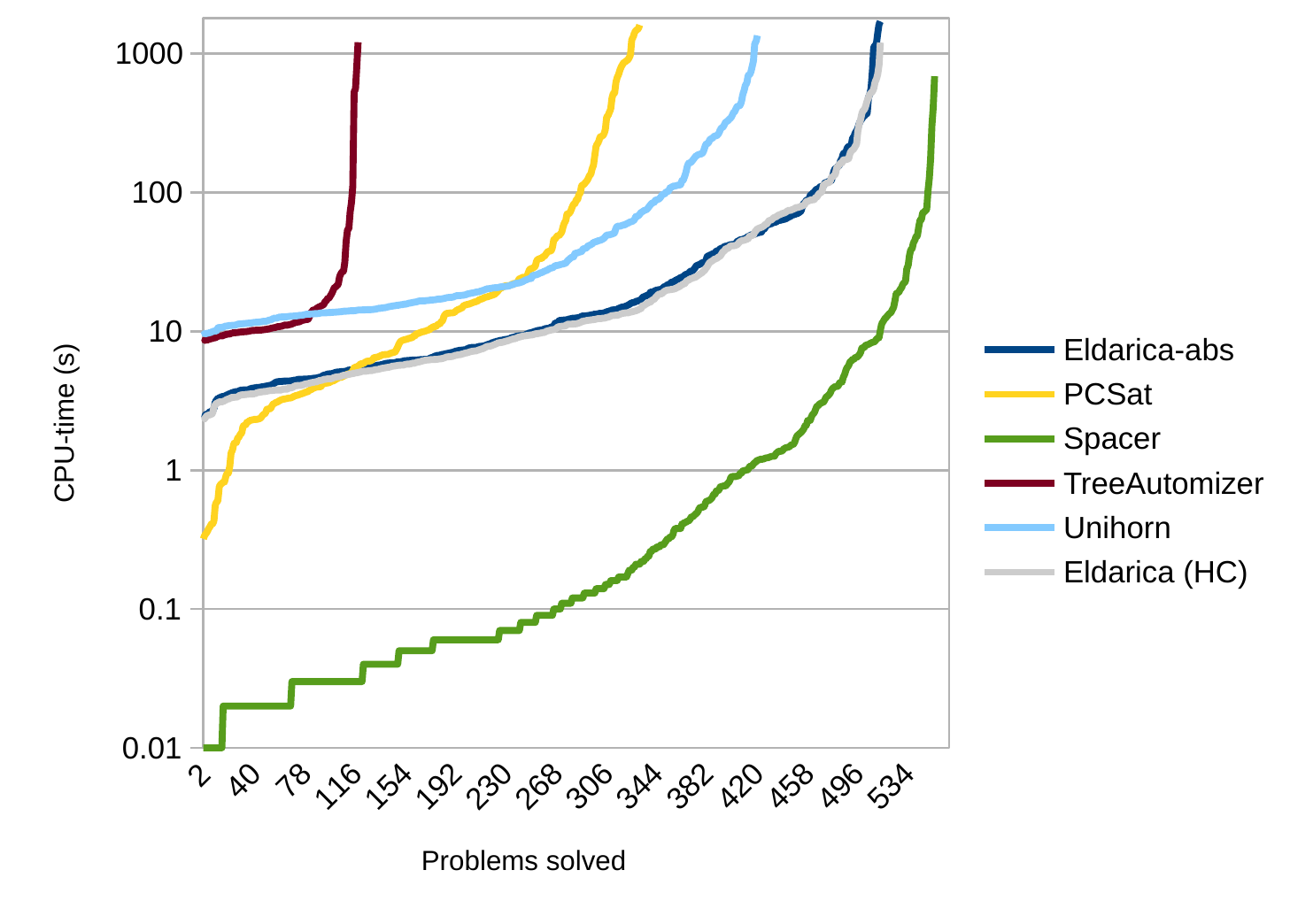}
  \includegraphics[scale=0.62]{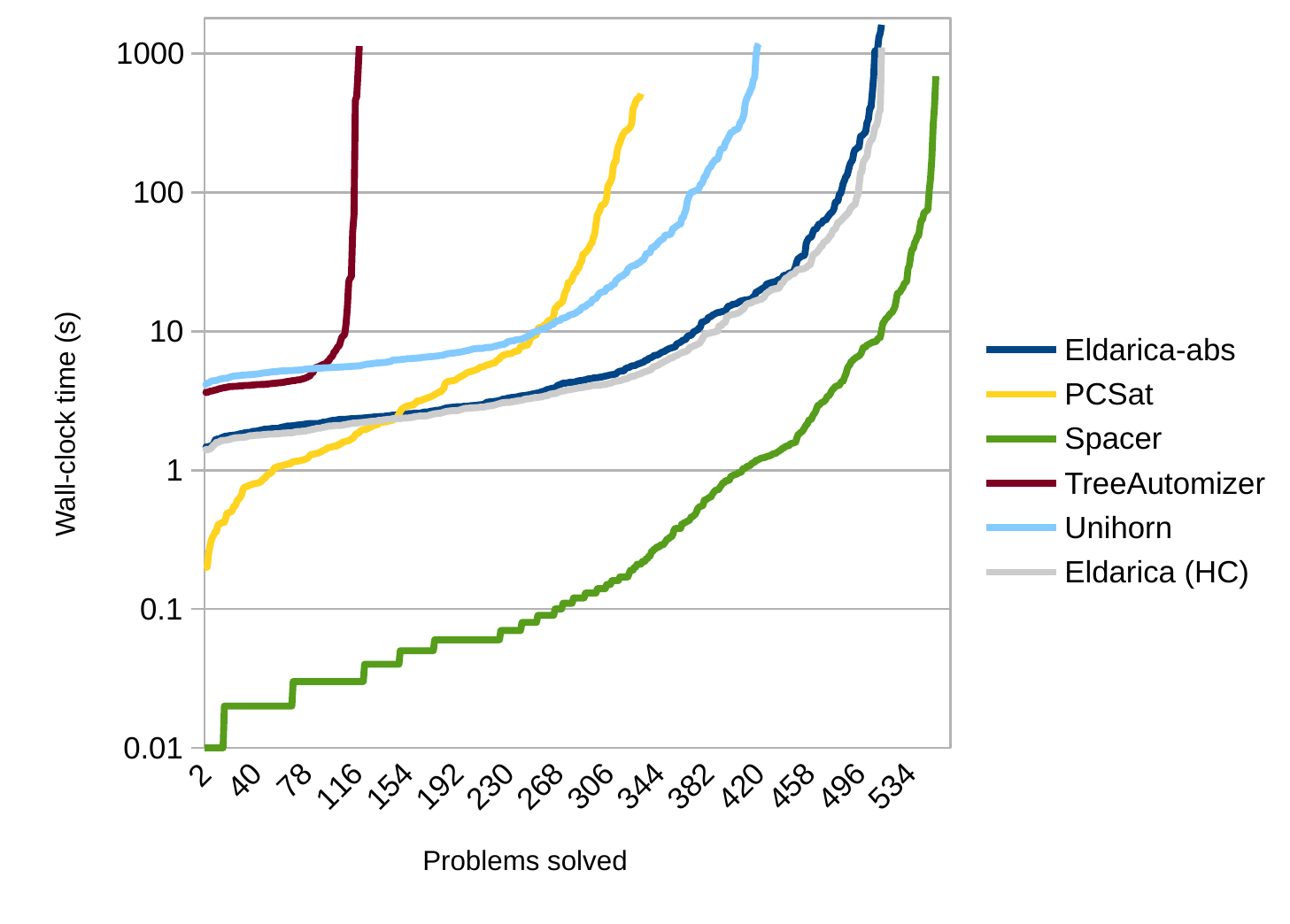}

  \caption{Solver performance on the 565 benchmarks of the \LIA track}
  \label{fig:results-LIA}
\end{figure}

\begin{figure}[p]
  \begin{center}
    \begin{tabular}{l*{7}{r}}
      \hline
      \textbf{Solver} & Score & \#sat & \#unsat & CPU time (s) &
            Wall-clock (s) & Speedup & SotAC
      \\\hline\hline
Spacer & 518 & 330 & 188 & 11.94 & 12.03 & 0.99 & 0.22\\\hline
Eldarica-abs & 477 & 300 & 177 & 57.26 & 39.59 & 1.45 & 0.20\\\hline
\cellcolor{lightgray}Eldarica (HC) & \cellcolor{lightgray}476 & \cellcolor{lightgray}300 & \cellcolor{lightgray}176 & \cellcolor{lightgray}48.58 & \cellcolor{lightgray}20.00 & \cellcolor{lightgray}2.43 & \cellcolor{lightgray}0.20\\\hline
U. Unihorn & 407 & 240 & 167 & 43.57 & 26.21 & 1.66 & 0.17\\\hline
IC3IA & 400 & 260 & 140 & 46.09 & 46.23 & 1.00 & 0.20\\\hline
PCSat & 329 & 191 & 138 & 37.91 & 12.23 & 3.10 & 0.17\\\hline
U. TreeAutomizer & 307 & 166 & 141 & 50.30 & 37.43 & 1.34 & 0.17\\\hline\hline
      Any solver & 558 & 356 & 202\\\hline
    \end{tabular}
  \end{center}

  \bigskip
  \includegraphics[scale=0.62,trim=15 0 100 0,clip]{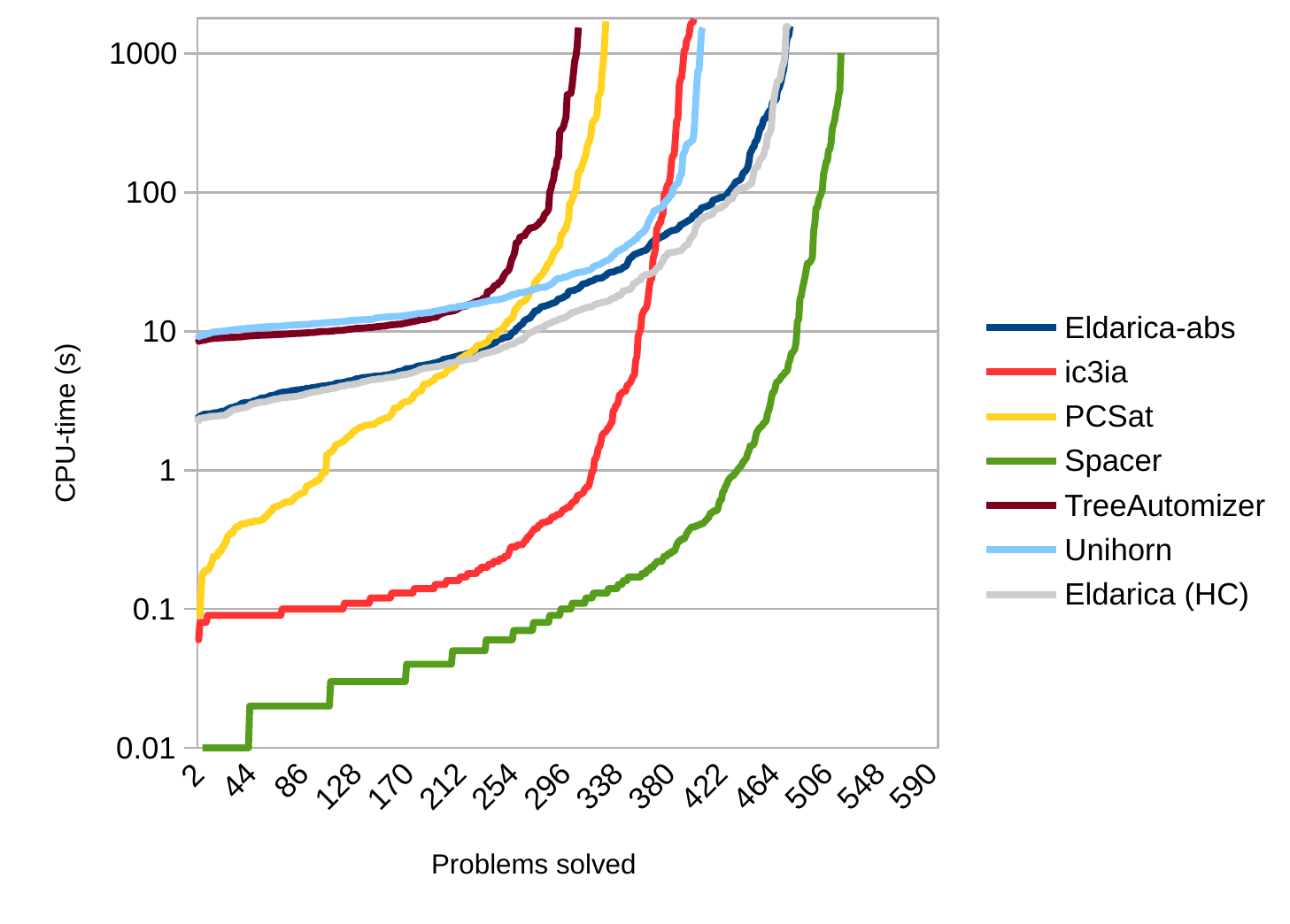}
  \includegraphics[scale=0.62]{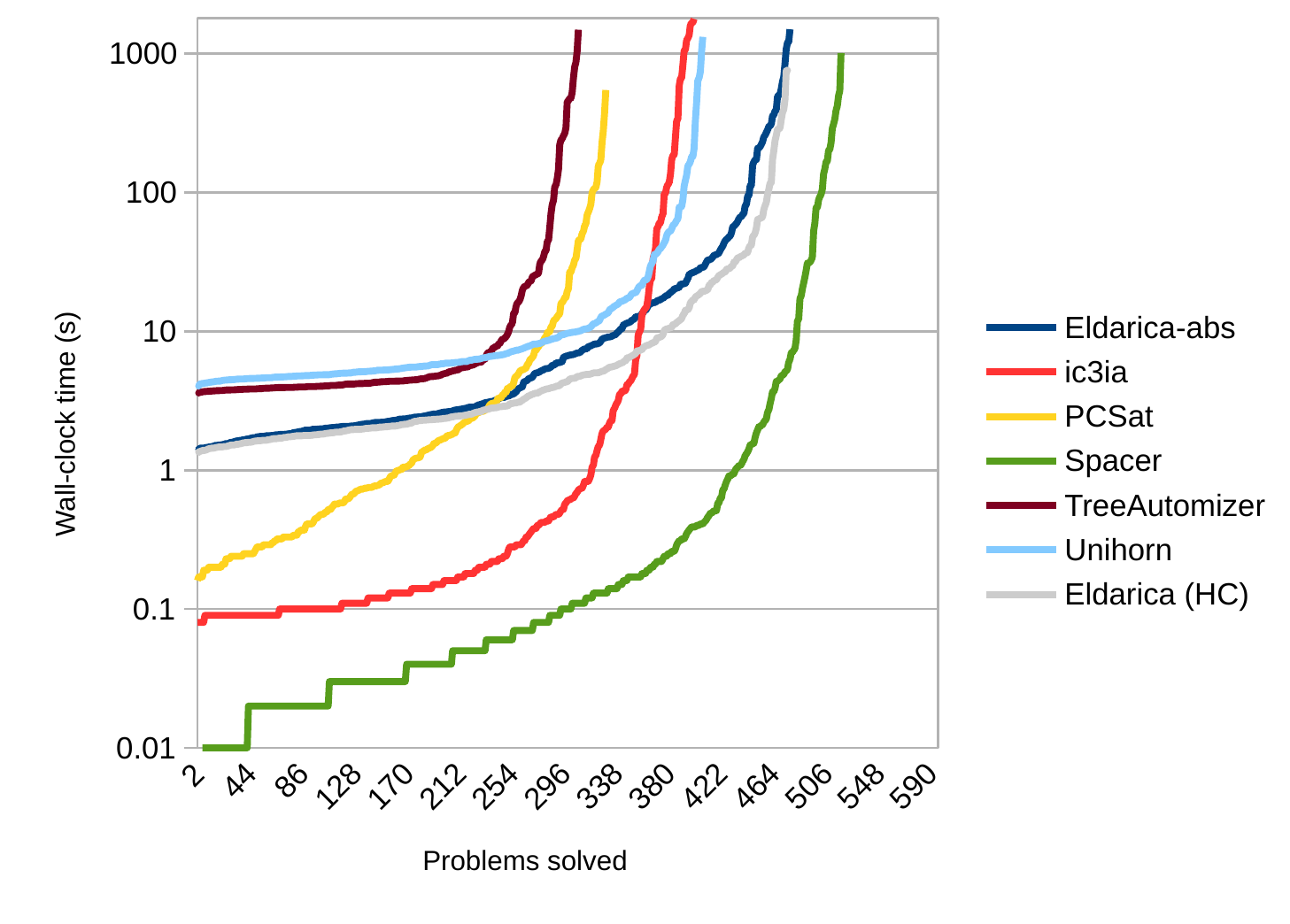}

  \caption{Solver performance on the 596 benchmarks of the \LIAlin track}
  \label{fig:results-LIAlin}
\end{figure}

\begin{figure}[p]
  \begin{center}
    \begin{tabular}{l*{7}{r}}
      \hline
      \textbf{Solver} & Score & \#sat & \#unsat & CPU time (s) &
            Wall-clock (s) & Speedup & SotAC
      \\\hline\hline
Spacer & 295 & 203 & 92 & 0.81 & 0.89 & 0.91 & 0.37\\\hline
U. Unihorn & 217 & 144 & 73 & 39.73 & 24.12 & 1.65 & 0.26\\\hline
ProphIC3 & 214 & 140 & 74 & 38.24 & 19.17 & 1.99 & 0.34\\\hline
IC3IA & 147 & 92 & 55 & 9.17 & 9.30 & 0.99 & 0.24\\\hline
U. TreeAutomizer & 147 & 100 & 47 & 31.49 & 21.46 & 1.47 & 0.22\\\hline
\cellcolor{lightgray}Eldarica (HC) & \cellcolor{lightgray}91 & \cellcolor{lightgray}91 & \cellcolor{lightgray}0 & \cellcolor{lightgray}106.80 & \cellcolor{lightgray}68.05 & \cellcolor{lightgray}1.57 & \cellcolor{lightgray}0.24\\\hline\hline
      Any solver & 350 & 250 & 100\\\hline
    \end{tabular}
  \end{center}

  \bigskip
  \includegraphics[scale=0.62,trim=15 0 100 0,clip]{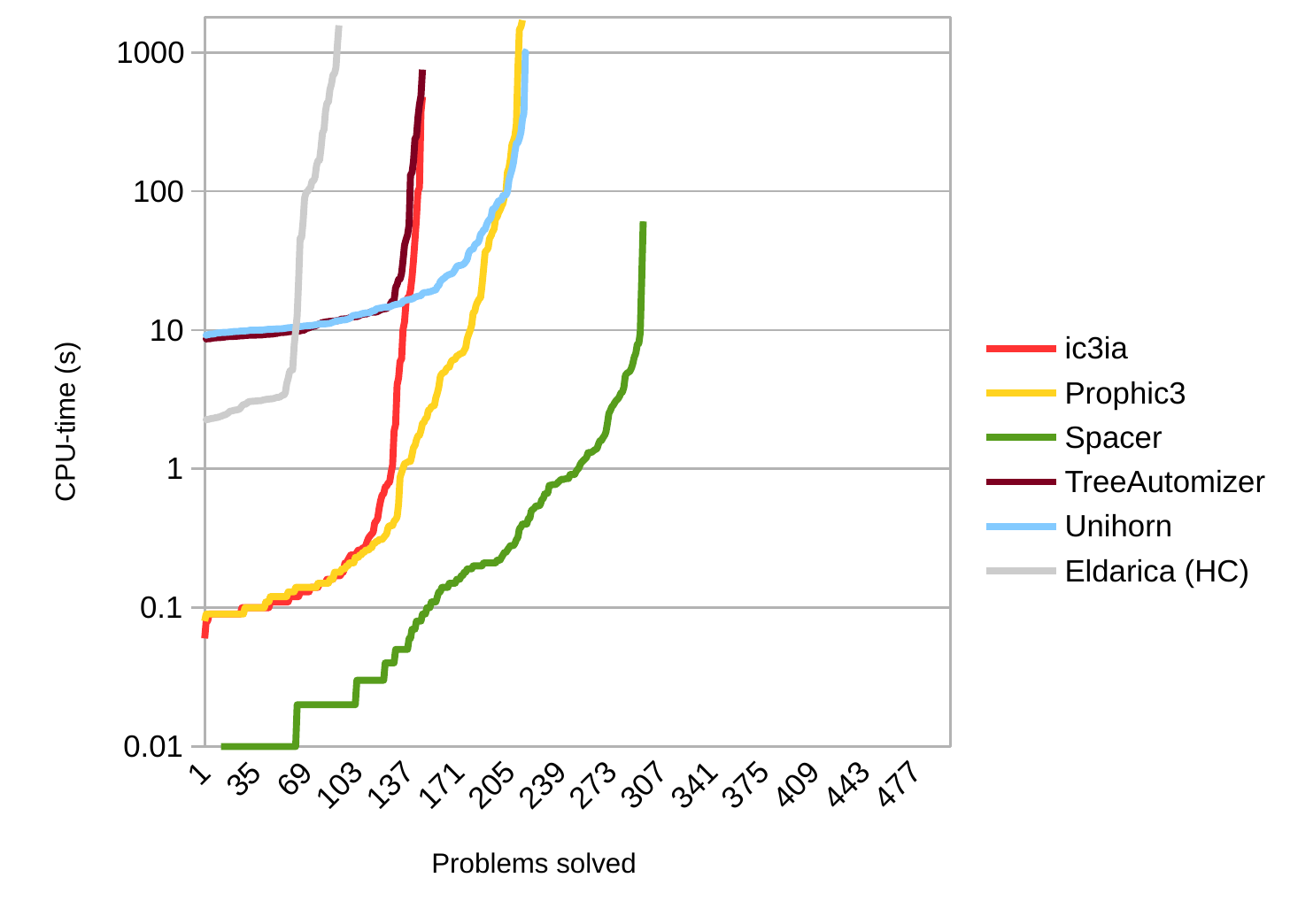}
  \includegraphics[scale=0.62]{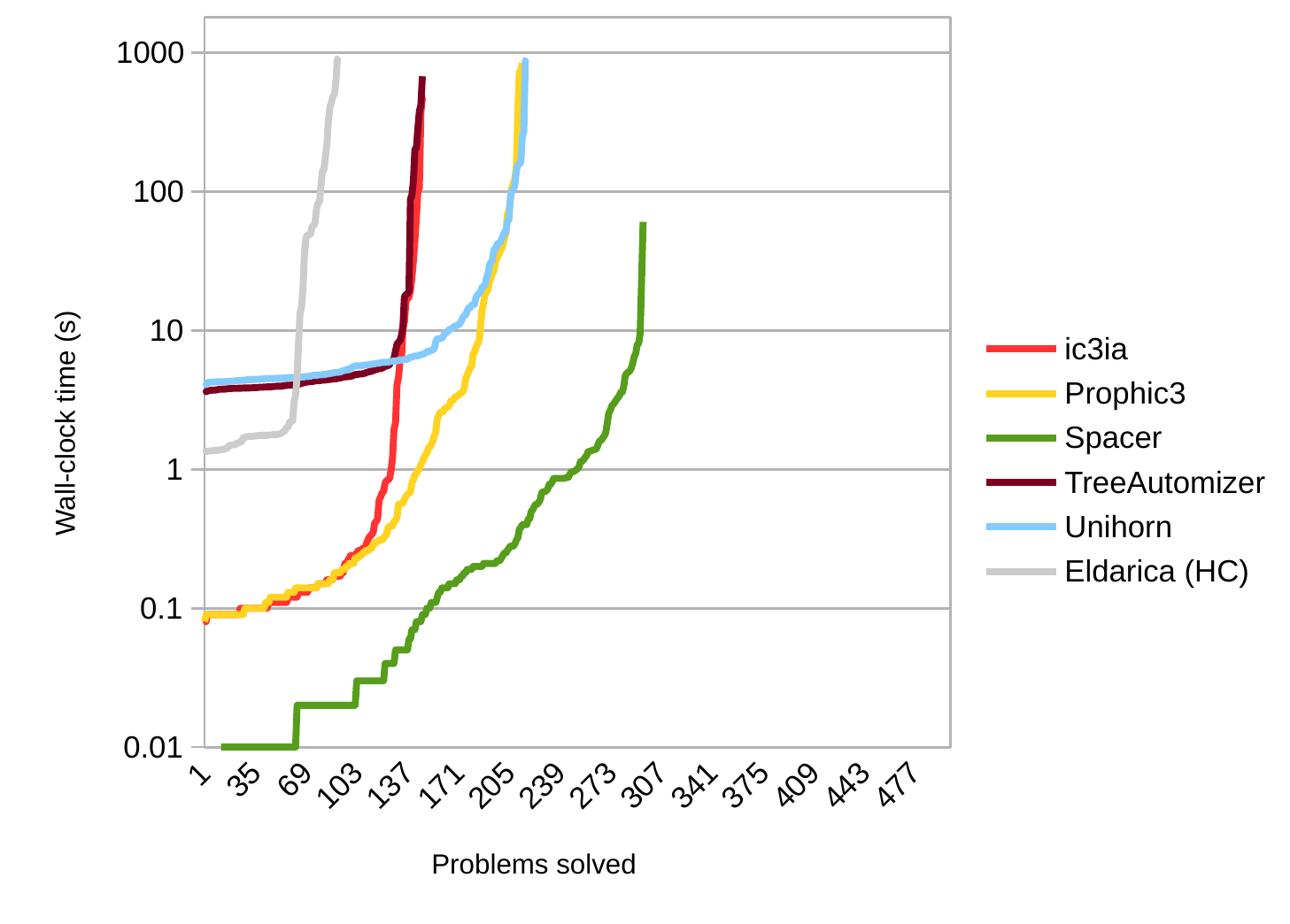}

  \caption{Solver performance on 500 benchmarks of the \LIAlinar track
    (one benchmark on which Spacer and Ultimate Unihorn give
    conflicting answers is not counted)}
  \label{fig:results-LIAlinar}
\end{figure}

\begin{figure}[p]
  \begin{center}
    \begin{tabular}{l*{7}{r}}
      \hline
      \textbf{Solver} & Score & \#sat & \#unsat & CPU time (s) &
            Wall-clock (s) & Speedup & SotAC
      \\\hline\hline
IC3IA & 468 & 378 & 90 & 136.94 & 137.05 & 1.00 & 0.29\\\hline
Sally-parallel & 439 & 360 & 79 & 138.81 & 47.37 & 2.93 & 0.24\\\hline
Sally-decomposing-itp & 438 & 357 & 81 & 107.61 & 107.68 & 1.00 & 0.24\\\hline
Spacer & 346 & 270 & 76 & 176.75 & 176.86 & 1.00 & 0.22\\\hline
U. TreeAutomizer & 168 & 131 & 37 & 239.75 & 202.11 & 1.19 & 0.19\\\hline
U. Unihorn & 160 & 103 & 57 & 213.33 & 158.57 & 1.35 & 0.18\\\hline\hline
      Any solver & 481 & 388 & 93\\\hline
    \end{tabular}
  \end{center}

  \bigskip
  \includegraphics[scale=0.62,trim=15 0 110 0,clip]{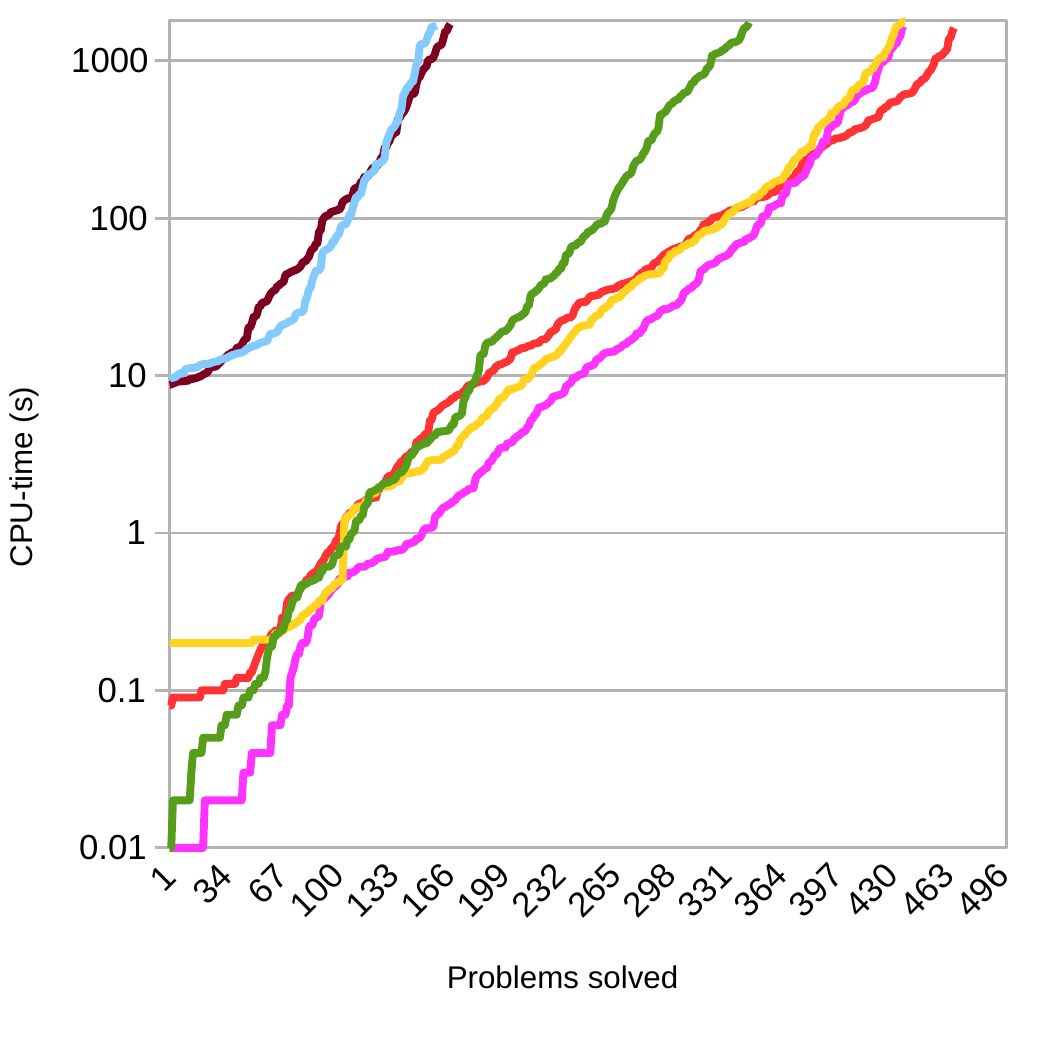}
  \includegraphics[scale=0.62]{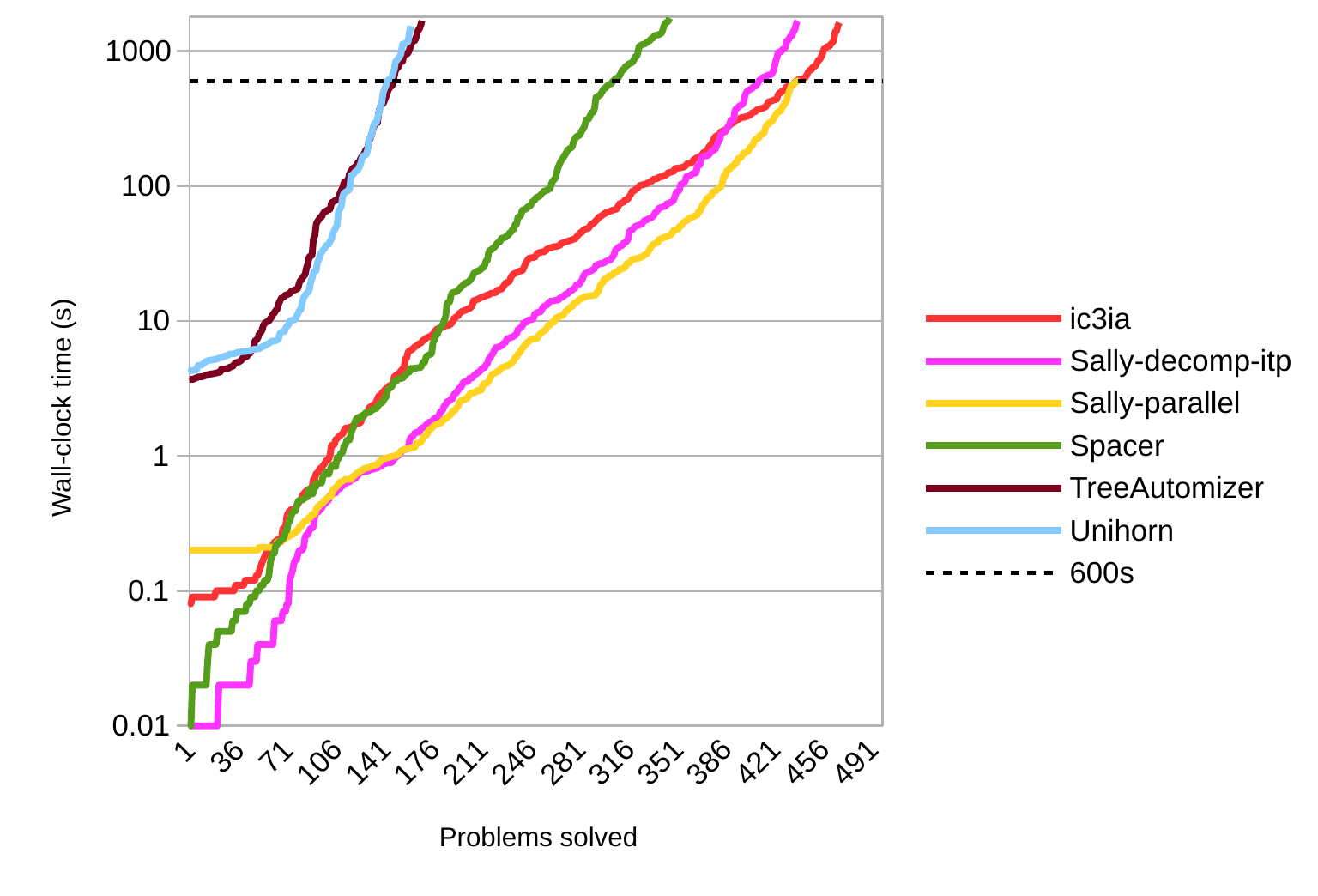}

  \caption{Solver performance on the 499 benchmarks of the \LRATS track}
  \label{fig:results-LRA-TS}
\end{figure}

Detailed results for the four tracks can be found in
Figures~\ref{fig:results-LIA}, \ref{fig:results-LIAlin},
\ref{fig:results-LIAlinar}, and \ref{fig:results-LRA-TS}.

\subsection{Observed Inconsistencies in the Competition Runs, and
  Fixes}

\paragraph{\LIAlinar:}

Only one case of inconsistent results was observed in the competition
runs, namely the results for benchmark
\verb!chc-LIA-lin-arrays_381.smt2! in the \LIAlinar track. Ultimate
Unihorn claimed that this benchmark is satisfiable, whereas
Spacer reported that it is unsatisfiable; all other solvers timed out.

This issue was discussed with the authors of the solvers. Spacer can
produce an internal counterexample for the problem, but no
counterexample for the complete input problem that could be verified
by independent tools. Ultimate Unihorn does not have
functionality to output models. No obvious bug was found in either of
the tools. This means that there is no immediate way to establish the
actual status of the benchmark beyond doubt, and no way to tell which
of the solvers gave the right answer; but clearly one of the tools
contains a bug.

Since no further inconsistencies were observed in the competition, in
this particular case the organiser decided to remove the benchmark
\verb!chc-LIA-lin-arrays_381.smt2! and not count Spacer's nor Ultimate
Unihorn's answer. The issue highlights a problem in the
competition design, however, which should be addressed in the next
competition (see Section~\ref{sec:next} for more thoughts about this).

\paragraph{Fixes in IC3IA and ProphIC3:}

After the submission deadline, the team submitting the solvers IC3IA
and ProphIC3 reported that they had found a bug affecting the
soundness of both tools in the \LIAlinar track, and provided new
versions of the  tools. No inconsistencies were observed for either
tool in the competition runs, but since there was enough time the
experiments in \LIAlinar were repeated with the new versions of the
solvers. This changed the results in \LIAlinar in the following way:
\begin{center}
  \begin{tabular}{l*{2}{@{\qquad}c}}
    \hline
    & \#sat & \#unsat
    \\\hline\hline
    IC3IA (original) & 96 & 47
    \\\hline
    IC3IA (fixed) &  92 & 55
    \\\hline
    ProphIC3 (original) & 183 & 74
    \\\hline
    ProphIC3 (fixed) & 140 & 74
    \\\hline
  \end{tabular}
\end{center}
The performance of IC3IA increased slightly, and the new version
reached the same score as Ultimate TreeAutomizer. The performance of
ProphIC3, in contrast, decreased significantly, and ProphIC3 moved
from the second to the third place.

\subsection{Resource Budgets: CPU time vs.\ Wall-clock time}

In the preparation phase of \CHCC, there was a discussion with several
teams about the use of CPU time vs.\ wall-clock time to define the
resource budget of solvers, and the possibility to have separate
tracks in future editions of \CHC for parallel solvers (with
wall-clock time budget and no limit on CPU time). For \CHCC, such
tracks were not introduced in the end, and the primary resource limit
was the available CPU time, but adding tracks for parallel solvers can
be an interesting new feature for \CHCCnew, or beyond.

It is meaningful to analyse the \CHCC outcomes in this light. An
immediate observation is that the three LIA tracks show very different
behaviour with respect to runtime than the \LRATS track. In the LIA
tracks, the average CPU time of solvers is usually significantly below
100s, and the cactus plots show that the ranking of solvers hardly
changes above 100s CPU time (Figures~\ref{fig:results-LIA},
\ref{fig:results-LIAlin}, \ref{fig:results-LIAlinar}). Only few
benchmarks in those tracks can be solved with CPU time above
600s. This means that CPU time (and therefore also wall-clock time) is
essentially not a limiting factor in the LIA tracks, the current
solvers already hardly utilise the available computation time. Since
there is plenty of computation time available, already at this point
solvers can use portfolios and run different configurations in
parallel, as done by some of the participating solvers. To make
specific parallel LIA tracks interesting, it would be necessary to
change the scoring scheme of \CHC: no longer count just the number
of solved benchmarks, but also factor in the required wall-clock time
in the ranking of solvers.

The situation is different in \LRATS: in this track the average CPU
time used by solvers is between 100s and 300s, and the cactus plots
show interesting developments even after 600s CPU time. Since the
maximum speed-up observed in \LRATS is 2.93, the cactus plot for
wall-clock performance should be interpreted only up to a time limit
of around 600s, beyond 600s the results are limited by the available
CPU time. Limiting the wall-clock time to 600s would indeed change the
ranking of solvers, with the parallel computation used in pSally
paying off, and other solvers could of course be optimised in a
similar way. It would be an interesting experiment to repeat the
evaluation of \LRATS with unlimited CPU time, and wall-clock time
limited to 1800s, which would amount to a parallel track.

In summary, for the next editions of \CHC parallel tracks are
mainly interesting for \LRATS, where solvers can indeed utilise the
available computation time. In the LIA tracks the solvers are mainly
limited by the implemented algorithms and heuristics, and less by the
available computation time.

\section{Conclusions}
\label{sec:next}

The organiser would like to congratulate the winners of the four \CHCC
tracks, Spacer and IC3IA, and all solvers and people submitting
solvers for the excellent performance this year! Thanks go to all
people who have been helping with infrastructure, scripts, benchmarks,
or in other ways, see the acknowledgements in the introduction; and to
the HCVS workshop for hosting \CHC!

\bigskip
\noindent
In order to keep \CHC an interesting and relevant competition, the
organiser also identified several questions and issues that should be
discussed and addressed in the next editions:
\begin{itemize}
\item Duplicate benchmarks included in multiple repositories. The
  selection of benchmarks in \CHCC probably contained some benchmarks
  repeatedly, for instance benchmarks from the repositories sv-comp or
  chc-comp19-benchmarks that were pre-processed using different tools
  initially, and thus not identified as duplicates. The outcome of the
  competition was probably not affected very much by this, but this is
  an issue that should be addressed in the long run. To keep the
  decentralised model of storing benchmarks in \CHC, one could
  maintain a global list of all available unique benchmarks.
\item A more systematic method for hardness rating of benchmarks is
  needed. Some rating is required to select interesting benchmarks,
  but any rating has the potential of changing the outcome of the
  competition. The use of Eldarica and Ultimate Unihorn (in
  the versions from \CHCCold) for problem rating possibly puts the
  \CHCC versions of those solvers at a disadvantage (or advantage)
  compared to other solvers not used for rating. For instance, if
  solvers~A and B show completely uncorrelated performance on some set
  of benchmarks, then picking 50\% easy and 50\% hard benchmarks for
  solver~A will not affect the expected outcome for solver~B, but will
  fix the outcome of solver~A to 50\%.
\item A better way has to be found to handle cases of inconsistent
  results, as observed this year in the \LIAlinar track. There are
  different solutions: (i)~only use benchmarks with known status in
  the competition; (ii)~require solvers to produce, on demand, models
  or counterexamples when they claim to be able to solve a
  problem. This requires a discussion.
\item An approach to determine and store the expected outcome of the
  individual benchmarks.
\item A discussion is needed about new tracks to be added in the
  competition: parallel tracks, or tracks with new background
  theories, for instance algebraic data-types or bit-vectors?
\item \textbf{A bigger set of benchmarks is needed, and all users and
    tool authors are encouraged to submit benchmarks!} In particular,
  in the \LIA and \LRATS tracks, the competition results indicate that
  harder benchmarks are required.
\end{itemize}

\newpage
\section{Solver Descriptions}
\label{sec:solvers}

The tool descriptions in this section were contributed by the tool
submitters, and the copyright on the texts remains with the individual
authors.

\makeatletter
\renewcommand\paragraph{\@startsection{paragraph}{4}{\z@}%
                                    {2ex \@plus1ex \@minus.2ex}%
                                    {-1em}%
                                    {\normalfont\normalsize\bfseries}}
\makeatother

\newcommand{\toolname}[1]{\bigskip\section*{#1}}
\newcommand{\toolsubmitter}[2]{%
  \noindent
  \begin{tabular}[t]{@{}l@{}}
    #1\\#2
  \end{tabular}\hspace*{8ex}}
\newcommand{\toolalgorithm}{\paragraph{Algorithm.}}
\newcommand{\toolarchitecture}{\paragraph{Architecture and Implementation.}}
\newcommand{\toolconfiguration}{\paragraph{Configuration in \CHCC.}}
\newcommand{\toollink}[2]{\par\medskip\noindent\url{#1}\\#2}




\toolname{Eldarica-abs}


\toolsubmitter{Xiaozhen Zhang}{Dalian University of Technology, China}
\toolsubmitter{Weiqiang Kong}{Dalian University of Technology, China}

\toolalgorithm

Eldarica-abs is a variant of Eldarica \cite{FMCAD2018HojjatRummer},
which is a model checker for Horn clauses, Numerical Transition
Systems, and software programs. Eldarica-abs mainly concentrates on
the modification of the exploration strategies of abstraction lattices
utilized in Eldarica. Identical with Eldarica, the inputs of
Eldarica-abs can be read in a variety of formats, including SMT-LIB v2
and Prolog for Horn clauses, and fragments of Scala and C for software
programs; and these inputs can be analysed using a variant of the
Counterexample-Guided Abstraction Refinement (CEGAR) method, in which
interpolation-based techniques are used to synthesis new predicates
for CEGAR. Different from Eldarica, Eldarica-abs takes the cost-guided
bidirectional heuristic search strategy to search constructed
abstraction lattices in order to compute the interpolants of good
quality \cite{ToAppearZhangKong}. As the goal elements are not only
maximal and feasible, but also of minimal cost, this search strategy
chooses from the top of the abstraction lattice to search downwards
and introduces the predecessors computing function to avoid the
repeated visit of the predecessors; and then, for the feasible
predecessors of one element under consideration, upward searching
strategy is taken to search an element whose all successors are all
infeasible; for the infeasible predecessors, the cost and
infeasibility information contained in these elements are utilized to
change the set of the elements to be searched; additionally, for the
selection of every element to be extended, we make the most of the
information contained in the visited elements set and the set of
elements to be visited, in order to choose the best candidate element
and then promote the performance of the searching process along an
effective path.

\toolarchitecture

The overall architecture of Eldarica-abs is the same as Eldarica and
it can be divided into three phases: input encoding phase;
preprocessing phase; CEGAR engine solving phase
\cite{FMCAD2018HojjatRummer}. Among these phases, interpolation
abstraction is introduced as an effective convergence heuristic
technique to compute the Craig interpolants of good quality, which are
utilized to inspire the acquirement of the right predicates necessary
in CEGAR process. In the course of obtaining suitable Craig
interpolants, Eldarica-abs takes a different cost-guided bidirectional
heuristic exploration strategy to search the constructed abstraction
lattices.

Based on the work of Eldarica, Eldarica-abs is also implemented in
Scala and depends on Java and Scala libraries and the Princess SMT
solver.

\toolconfiguration

Eldarica-abs was running with default options in the competition.

\toollink{https://github.com/zhangxiaozhen/Eldarica-abs}{BSD licence}


\newpage

\newcommand{\icia}{\textbf{ic3ia}\xspace}

\toolname{IC3IA 2020.05}


\toolsubmitter{Alberto Griggio}{Fondazione Bruno Kessler, Italy}
\toolsubmitter{Ahmed Irfan}{Stanford University, USA}
\toolsubmitter{Makai Mann}{Stanford University, USA}

\toolalgorithm

The tool is an open-source implementation of IC3 Modulo Theories via Implicit
Predicate Abstraction (IC3IA). It is one approach for extending IC3 to the
theory level, with the advantage that it can be applied to arbitrary theories
without theory-specific quantifier elimination procedures. 

\toolarchitecture

It depends on MathSAT 5.6.3~\cite{mathsat5}. The \icia code is distributed under
the GPLv3 license. This tool operates on transition systems rather than CHC. CHC
clauses are translated to Verification Modulo Theories (VMT) format using a
program distributed with \icia.

\toolconfiguration

The submitted solver is run using options \texttt{-solver-approx 1 -inc-ref 1}.

\toollink{https://es-static.fbk.eu/people/griggio/ic3ia/index.html}{GPLv3}




\toolname{PCSat}

\toolsubmitter{Yu Gu}{University of Tsukuba, Japan}
\toolsubmitter{Hiroshi Unno}{University of Tsukuba, Japan}

\toolalgorithm

PCSat is a solver for a general class of second-order constraints on predicate and function variables.  Its applications include but not limited to branching-time temporal verification, dependent refinement type inference, program synthesis, and infinite-state game solving.

PCSat is based on CounterExample-Guided Inductive Synthesis (CEGIS), with the support of multiple synthesis engines including template-based, decision-tree-based, and graphical-model-based~\cite{Unno2020} ones.

\toolarchitecture

PCSat is designed and implemented as a highly-configurable solver, allowing us to test various possible combinations of synthesis engines, example sampling strategies, and backend SAT/SMT solvers.  This design is enabled by the powerful module system and the metaprogramming features of the OCaml functional programming language.

\toolconfiguration

We adopted a parallel combination of the template-based and decision-tree-based synthesis engines.  Z3 and MiniSat are used as the backend SMT and SAT solvers, respectively.

\toollink{https://github.com/hiroshi-unno/coar}{Apache License 2.0}


\newpage


\newcommand{\prophic}{\textbf{prophic3}\xspace}


\toolname{ProphIC3}


\toolsubmitter{Makai Mann}{Stanford University, USA}
\toolsubmitter{Ahmed Irfan}{Stanford University, USA}
\toolsubmitter{Alberto Griggio}{Fondazione Bruno Kessler, Italy}

\medskip
\toolsubmitter{Oded Padon}{Stanford University, USA}
\toolsubmitter{Clark Barrett}{Stanford University, USA}

\toolalgorithm

The tool \prophic is a prototype implementation of a Counter-example Guided
Abstraction Refinement (CEGAR)~\cite{cegar} algorithm for model checking modulo
the theory of arrays. The algorithm abstracts the theory of arrays using
uninterpreted functions and lazily adds array axioms. Furthermore, it uses
counterexamples to add history and prophecy variables which can help find
simpler invariants, even reducing quantified invariants to quantifier-free
invariants in some cases. The approach wraps a standard model checker. The
underlying model checker must support all the theories used in the input
problem, except the theory of arrays.

\toolarchitecture

The \prophic prototype depends on \icia~\cite{ic3iacode}, an implementation of
IC3 Modulo Theories via Implicit Predicate Abstraction
(IC3IA)~\cite{ic3ia}. IC3IA is one approach for extending IC3~\cite{ic3}
to arbitrary theories. This is version 2020.05 of \icia. It depends on MathSAT
5.6.3~\cite{mathsat5}. This tool operates on transition systems rather than CHC.
CHC clauses are translated to Verification Modulo Theories (VMT) format using a
program distributed with \icia.

\toolconfiguration

The submitted solver is a portfolio approach which runs \prophic from git tag
\texttt{chccomp-2020} with option \texttt{-no-eq-uf}, as well as bounded model
checking~\cite{bmc} up to bound 100 on the concrete system.

\toollink{https://github.com/makaimann/prophic3}{GPLv3}




\newcommand{\psally}{pSally}
\newcommand{\sally}{Sally}


\toolname{\sally{} and \psally}


\toolsubmitter{Martin Blicha\footnotemark}{Universit\`{a} della Svizzera italiana, Switzerland}
\footnotetext{This submitted tool builds on the work of Dejan Jovanovi{\'c} and Bruno Dutertre. The tool and the related research were realised with significant contributions by various colleagues, in particular by Matteo Marescotti, Antti E.~J.~Hyv{\"a}rinen and Natasha Sharygina.}

\toolalgorithm

Our competition entry is derived from the model checker \sally{}~\cite{Jovanovic2016}, which we have enhanced in two ways: First, we have supplied our interpolating SMT solver OpenSMT with a specialized interpolation procedure for Linear Real Arithmetic (LRA) as part of \sally's backend. Secondly, we have implemented a parallel version of \sally{} (\psally) where multiple instances cooperate by sharing information discovered about the problem at hand. The current version of the tool is limited to solving safety of Transition Systems encoded in LRA.

The tool uses \sally{}'s PD-KIND engine~\cite{Jovanovic2016} as the core algorithm. PD-KIND strengthens the IC3 algorithm with $k$-induction and gradually builds a $k$-inductive safe invariant of the transition system.
It relies on an SMT solver for answering satisfiability queries, generalization and interpolation.
Our specialized interpolation procedure computes stronger interpolants than traditional interpolation algorithm and this has been shown to be useful in model checking scenarios~\cite{Blicha2019}. The parallel version on the other hand leverages a portfolio of interpolation algorithms for discovering useful facts about the system under analysis, as well as a cooperative framework for sharing the discovered information~\cite{Blicha2020}.

\toolarchitecture

The competition entry uses \sally{}'s PD-KIND reasoning engine, with the SMT solvers Yices2~\cite{Yices2014} and OpenSMT~\cite{OpenSMT2} for generalization and interpolation, respectively. OpenSMT uses an interpolation algorithm that computes \emph{decomposed} LRA interpolants~\cite{Blicha2019}.
The parallel version uses SMTS framework~\cite{smts18} for managing multiple instances and their communication.

\toolconfiguration \hfill\\
\noindent Command line options of \sally{}:\hfill\\
\texttt{
  \$ sally --engine pdkind --solver y2o2 --solver-logic
   QF\textunderscore{}LRA \hfill\\
   --pdkind-minimize-frames 
   --pdkind-minimize-interpolants  \hfill\\
   --opensmt2-itp-lra 4 --opensmt2-simplify\_itp 4 
   --yices2-mode dpllt --output-lang horn -i <input\textunderscore{}file>
}

\noindent Command line options of \psally{} for creating three instances and enabling the sharing of information:
\texttt{
\$ python3 smts.py -l -s3
}

\toollink{http://sri-csl.github.io/sally/}{GNU GENERAL PUBLIC LICENSE v2}
\toollink{https://zenodo.org/record/3484097}{MIT LICENSE}


\newcommand{\false}{\emph{false}\xspace}
\newcommand{\pob}{\textsc{pob}\xspace}
\newcommand{\Spacer}{\textsc{Spacer}\xspace}
\newcommand{\pobs}{\textsc{pob}s\xspace}

\toolname{\Spacer}


\toolsubmitter{Hari Govind V K}{University of Waterloo, Canada}
\toolsubmitter{Arie Gurfinkel}{University of Waterloo, Canada}

\toolalgorithm

\Spacer~\cite{DBLP:journals/fmsd/KomuravelliGC16} is an IC3/PDR-style algorithm
for solving linear and non linear CHCs. Given a set of CHCs, it iteratively
proves the unreachability of \false at larger and larger depths until a model is
found or the set of CHCs is proven unsatisfiable. To prove unreachability at a
particular depth, \Spacer recursively generates sets of predecessor
states~(called proof obligations~(\pobs)) from which \false can be derived and
blocks them. Once a \pob is blocked, \Spacer generalizes the proof to learn a
\emph{lemma} that blocks multiple \pobs. \Spacer uses many heuristics to learn
lemmas. These include interpolation, inductive generalization and quantifier
generalization. The latest version of Spacer presents a new heuristic for
learning lemmas~\cite{gspc}.

The current implementation of \Spacer supports linear and nonlinear CHCs in the
theory of Arrays, Linear Arithmetic and FixedSizeBitVectors. \Spacer can
generate both quantified and quantifier free models as well as resolution proof
of unsatisfiability.

\toolarchitecture
\Spacer is implemented on top of the \textsc{Z3} theorem prover. It uses many
SMT solvers implemented in \textsc{Z3}. Additionally, it implements a max-SMT
solver and an interpolating SMT solver.

\toolconfiguration


We used different configurations for different tracks. However, all the
preprocessing options and some \Spacer options remained the same in all the
tracks. Common configuration for all tracks:
\begin{verbatim}
fp.spacer.global=true fp.spacer.concretize=true fp.spacer.conjecture=true
fp.xform.tail_simplifier_pve=false fp.validate=true fp.spacer.mbqi=false 
\end{verbatim}
\vspace{0.1in}
Additionally, in the Arrays track, we used the quantifier generalization
strategies:
\begin{verbatim}
fp.spacer.q3.use_qgen=true fp.spacer.q3.instantiate=true fp.spacer.q3=true
\end{verbatim}
\vspace{0.1in}
In the LRA-TS track, we turned off interpolation:
\begin{verbatim}
fp.spacer.use_iuc=false
\end{verbatim}

\toollink{https://github.com/Z3Prover/z3}{MIT License}

\toolname{Ultimate TreeAutomizer 0.1.25-6b0a1c7}


\toolsubmitter{Matthias Heizmann}{University of Freiburg, Germany}
\toolsubmitter{Daniel Dietsch}{University of Freiburg, Germany}

\medskip
\toolsubmitter{Jochen Hoenicke}{University of Freiburg, Germany}
\toolsubmitter{Alexander Nutz}{University of Freiburg, Germany}

\medskip
\toolsubmitter{Andreas Podelski}{University of Freiburg, Germany}

\toolalgorithm

The \textsc{Ultimate TreeAutomizer} solver implements an approach that is based on tree automata~\cite{journals/corr/abs-1907-03998}.
In this approach potential counterexamples to satisfiability are considered as a regular set of trees.
In an iterative \nobreak{CEGAR} loop we analyze potential counterexamples.
Real counterexamples lead to an \textit{unsat} result.
Spurious counterexamples are generalized to a regular set of spurious counterexamples
and subtracted from the set of potential counterexamples that have to be considered.
In case we detected that all potential counterexamples are spurious, the result is \textit{sat}.
The generalization above is based on tree interpolation and
regular sets of trees are represented as tree automata.


\toolarchitecture

\textsc{TreeAutomizer} is a toolchain in the 
\textsc{Ultimate} framework.
This toolchain first parses the CHC input and then runs the \texttt{treeautomizer} plugin which
implements the above mentioned algorithm.
We obtain tree interpolants from the SMT solver SMTInterpol%
\footnote{\url{https://ultimate.informatik.uni-freiburg.de/smtinterpol/}}%
~\cite{cade/HoenickeS18}.
For checking satisfiability, we use the
 Z3 SMT solver%
\footnote{\url{https://github.com/Z3Prover/z3}}%
.
The tree automata are implemented in \textsc{Ultimate}'s automata library%
\footnote{\url{https://ultimate.informatik.uni-freiburg.de/automata_library}}%
.
The \textsc{Ultimate} framework is written in Java and build upon the Eclipse Rich Client Platform (RCP). The source code is available at
GitHub\footnote{\url{https://github.com/ultimate-pa/}}.


\toolconfiguration

Our StarExec archive for the competition is shipped with the \texttt{bin/starexec\_run\_default}
shell script calls the \textsc{Ultimate} command line interface with the\linebreak
\texttt{TreeAutomizer.xml} toolchain file and
the \texttt{TreeAutomizerHopcroftMinimization.epf} settings file.
Both files can be found in toolchain (resp. settings) folder of \textsc{Ultimate}'s repository.


\toollink{https://ultimate.informatik.uni-freiburg.de/}{LGPLv3 with a linking exception for Eclipse RCP}

\toolname{Ultimate Unihorn 0.1.25-6b0a1c7}


\toolsubmitter{Matthias Heizmann}{University of Freiburg, Germany}
\toolsubmitter{Daniel Dietsch}{University of Freiburg, Germany}

\medskip
\toolsubmitter{Jochen Hoenicke}{University of Freiburg, Germany}
\toolsubmitter{Alexander Nutz}{University of Freiburg, Germany}

\medskip
\toolsubmitter{Andreas Podelski}{University of Freiburg, Germany}

\toolalgorithm

\textsc{Ultimate Unihorn} reduces the satisfiability problem for a set of constraint Horn clauses
to a software verfication problem.
In a first step \textsc{Unihorn} applies a 
yet unpublished translation in which the constraint Horn clauses
are translated into a recursive program
that is nondeterministic and
whose correctness is specified by an assert statement
The program is correct (i.e., no execution violates the assert statement)
if and only if the set of CHCs is satisfiable.
For checking whether the recursive program satisfies its specification,
Unihorn uses \textsc{Ultimate Automizer}~\cite{tacas/HeizmannCDGHLNM18}
which implements an automata-based approach to software verification~\cite{cav/HeizmannHP13}.


\toolarchitecture

\textsc{Ultimate Unihorn} is a toolchain in the 
\textsc{Ultimate} framework.
This toolchain first parses the CHC input and then runs the \texttt{chctoboogie} plugin which
does the translation from CHCs into a recursive program.
We use the Boogie
language to represent that program.
Afterwards the default toolchain for verifying a recursive Boogie programs by \textsc{Ultimate Automizer} is applied.
The \textsc{Ultimate} framework shares the libraries for handling SMT formulas with the SMTInterpol SMT solver.
While verifying a program, \textsc{Ultimate Automizer} needs SMT solvers
for checking satisfiability,
for computing Craig interpolants and
for computing unsatisfiable cores.
The version of \textsc{Unihorn} that participated in the competition
used the SMT solvers SMTInterpol%
\footnote{\url{https://ultimate.informatik.uni-freiburg.de/smtinterpol/}}%
and Z3%
\footnote{\url{https://github.com/Z3Prover/z3}}%
.
The \textsc{Ultimate} framework is written in Java and build upon the Eclipse Rich Client Platform (RCP). The source code is available at
GitHub\footnote{\url{https://github.com/ultimate-pa/}}.


\toolconfiguration

Our StarExec archive for the competition is shipped with the \texttt{bin/starexec\_run\_default}
shell script calls the \textsc{Ultimate} command line interface with the\linebreak
\texttt{AutomizerCHC.xml} toolchain file and
the \texttt{AutomizerCHC\_No\_Goto.epf} settings file.
Both files can be found in toolchain (resp. settings) folder of \textsc{Ultimate}'s repository.


\toollink{https://ultimate.informatik.uni-freiburg.de/}{LGPLv3 with a linking exception for Eclipse RCP}

\newpage

\toolname{Eldarica v2.0.3 (Hors Concours)}


\toolsubmitter{Hossein Hojjat}{University of Tehran, Iran}
\toolsubmitter{Philipp R\"ummer}{Uppsala University, Sweden}

\toolalgorithm

Eldarica~\cite{FMCAD2018HojjatRummer} is a Horn solver applying
classical algorithms from model checking: predicate abstraction and
counterexample-guided abstraction refinement (CEGAR).  Eldarica can
solve Horn clauses over linear integer arithmetic, arrays, algebraic
data-types, and bit-vectors.  It can process Horn clauses and programs
in a variety of formats, implements sophisticated algorithms to solve
tricky systems of clauses without diverging, and offers an elegant API
for programmatic use.

\toolarchitecture

Eldarica is entirely implemented in Scala, and only depends on Java or
Scala libraries, which implies that Eldarica can be used on any
platform with a JVM. For computing abstractions of systems of Horn
clauses and inferring new predicates, Eldarica invokes the SMT solver
Princess~\cite{princess08} as a library.

\toolconfiguration

Eldarica is in the competition run with the option \verb!-abstractPO!,
which enables a simple portfolio mode: two instances of the solver are
run in parallel, one with the default options, and one with the option
\verb!-abstract:off! to switch off the interpolation abstract
technique.

\toollink{https://github.com/uuverifiers/eldarica}{BSD licence}


\newpage
\bibliographystyle{eptcs}
\bibliography{refs}

\end{document}